\documentclass[epj,twocolumn]{webofc}
\usepackage[varg]{txfonts}   
\usepackage{graphicx}

\newcommand{\teff}{$T_{\rm eff}$}

\newcommand{\feh}{[Fe/H]}
\newcommand{\mh}{[M/H]}
\newcommand{\logg}{$\log g$}
\newcommand{\mlsep}{$\langle \Delta \nu \rangle$}
\newcommand{\numax}{$\nu_{\rm max}$}

\newcommand{\msol}{M$_{\odot}$}
\newcommand{\rsol}{R$_{\odot}$}
\newcommand{\lsol}{L$_{\odot}$}

\wocname{epj}
\woctitle{Seismology of the Sun and the Distant Stars 2016}
\begin{document}
\title{Seismic inference of 57 stars using full-length \emph{Kepler} data sets}
%

\author{\firstname{Orlagh} \lastname{Creevey}\inst{1}\fnsep\thanks{\email{ocreevey@oca.eu}} \and
        \firstname{Travis~S.} \lastname{Metcalfe}\inst{2,3} \and
        \firstname{David} \lastname{Salabert}\inst{4}\fnsep{\thanks{Visiting Scientist at 1}} \and
        \firstname{Savita} \lastname{Mathur}\inst{2} \and
        \firstname{Mathias} \lastname{Schultheis}\inst{1} \and
        \firstname{Rafael~A.} \lastname{Garc\'ia}\inst{4} \and
        \firstname{Fr\'ed\'eric} \lastname{Th\'evenin}\inst{1} \and
        \firstname{Micha\"el} \lastname{Bazot}\inst{5} \and
        \firstname{Haiying} \lastname{Xu}\inst{6}
}

\institute{Universit\'e C\^ote d'Azur, 
Laboratoire Lagrange, 
   CNRS, Blvd de l'Observatoire, 
  CS 34229, 06304 Nice cedex 4, France
\and
Center for Extrasolar Planetary Systems, Space Science Institute, 4750 Walnut St.\ Suite 205, Boulder CO 80301, USA
\and
Visiting Scientist, National Solar Observatory, 3665 Discovery Dr., 
Boulder CO 80303, USA
\and
Laboratoire AIM, CEA/DRF-CNRS, Universit\'e Paris 7 Diderot, 
IRFU/SAp, Centre de Saclay, 91191 Gif-sur-Yvette, France
\and
Center for Space Science, NYUAD Institute, New York University Abu Dhabi, PO Box 129188, Abu Dhabi, UAE
\and
Computational \& Information Systems Laboratory, NCAR, P.O. Box 3000, Boulder CO 80307, USA
          }

\abstract{%
  We present stellar properties of 57 stars from a seismic inference using full-length data sets from \emph{Kepler} (mass, age, radius, distances). These stars comprise active stars, planet-hosts, solar-analogs, and binary systems. 
We validate the distances derived from the astrometric Gaia-Tycho solution.  Ensemble analysis of the stellar properties reveals a trend of mixing-length parameter with the surface gravity and effective temperature. We derive a linear relationship with the seismic quantity $\langle r_{02} \rangle$ to estimate the stellar age. Finally, we define the stellar regimes where the Kjeldsen et al (2008) empirical surface correction for 1D model frequencies is valid.
}
\maketitle%
\section{Introduction}
\label{intro}
Time-series analysis of the full-length \emph{Kepler} data sets of solar-like 
main sequence and subgiants stars is presented 
in \citep{Lund2016}.  They identify modes of oscillations for 66 stars.
Using the {\it Asteroseismic Modeling Portal}, AMP\footnote{https://amp.phys.au.dk/},
we analyse the individual frequency data from \citep{Lund2016} for 57 of these
stars and we supplement them with spectroscopic data from 
\citep{Ramirez2009,Buchhave2012,Pinsonneault2012,Huber2013,Chaplin2014,Pinsonneault2014}. 
In this version of AMP, AMP~1.3, we fit the 
frequency separation ratios $r_{01}$ and $r_{02}$ \citep{Roxburgh2003}
to determine the optimal models for each star in our sample. 

The optimization method in AMP is genetic algorithm (GA) which 
efficiently samples the full parameter ranges without imposing 
constraints on unknown parameters, such as the mixing-length parameter
or the initial chemical composition.  The results of the GA
is a dense clustering of models (1000s) around the optimal parameters.
We analyse the distribution of these models to determine the stellar
properties, such as mass, radius and age, along with their uncertainties.
Details of the methods can be found in \citep{Metcalfe2003,Metcalfe2009,Creevey2016}, with the
most recent reference containing the tables of stellar parameters for the 
sample of 57 stars presented here.  The results are 
validated using solar data and independent determinations of luminosity, radii, and ages.

In these proceedings we analyse the derived stellar properties of our sample. 
We compare their predicted distances
with the recent solution provided by the Gaia-Tycho analysis (TGAS, \citep{tgas2015}, 
Sect.~\ref{sec:distances}).  
In Section~\ref{sec:stellartrends} we derive expressions for estimating
the mixing-length parameter and the stellar age based on observed properties.
Finally, in Section~\ref{sec:surface} 
we explore the range of parameters where the correction proposed 
by \citep{Kjeldsen2008} to mimic the so-called {\it surface effect} related
to seismic data is useful.

\begin{figure}[h]
\centering
\includegraphics[width=0.45\textwidth,clip]{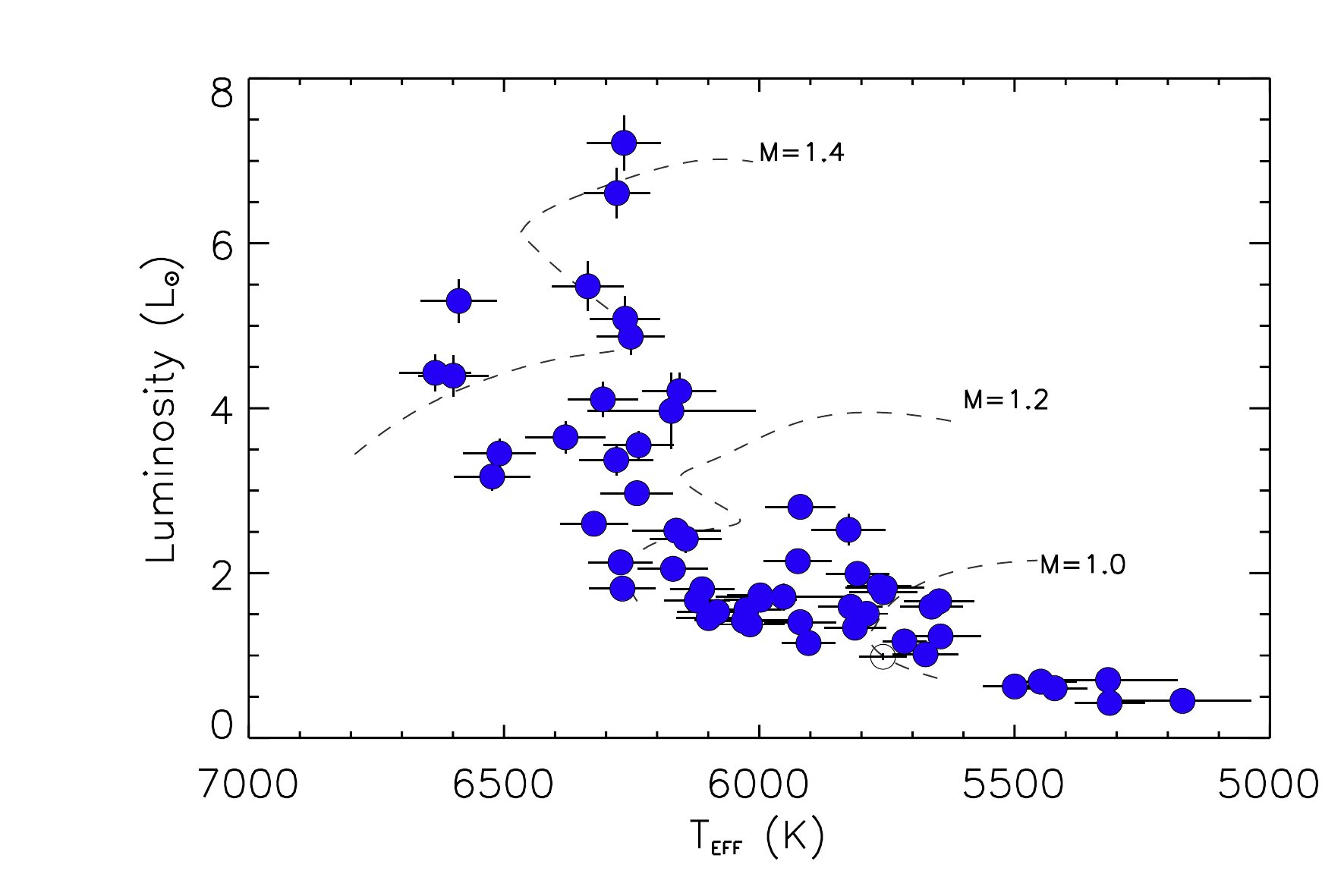}
    \caption{\label{fig:hrdiag}HR diagram showing the position of the 57 \emph{Kepler} stars 
used in this work.  
    Evolutionary tracks for solar-metallicity models with 1.0, 1.2, and 1.4 \msol\ stellar masses are shown.}
\end{figure}

\section{Asteroseismic distances}
\label{sec:distances}
We use the stellar luminosity, $L$, constrained from the asteroseismic
analysis to compute the stellar distance, as a parallax.
The model surface gravity and the observed
\teff\ and \feh\ are used to derive 
the amount of interstellar absorption between the top
of the Earth's atmosphere and the star, $A_{Ks}$, by applying 
the isochrone method described in \citep{schultheis2014}.  
Here, ths subscript $Ks$ refers to the 2MASS $K_s$ filter \citep{2MASS}.
We compute the  
bolometric correction $BC_{Ks}$, using 
$BC_{Ks} = 4.514650 -0.000524 T_{\rm eff}$ \citep{marigo2008} 
where the solar bolometric magnitude is 4.72 mag.
With $L$, $K_s$, $A_{Ks}$, and $BC_{Ks}$ we derive the 
the distance to each of the stars
in our sample, and then its parallax.

\begin{figure}
\centering
\includegraphics[width=0.45\textwidth,clip]{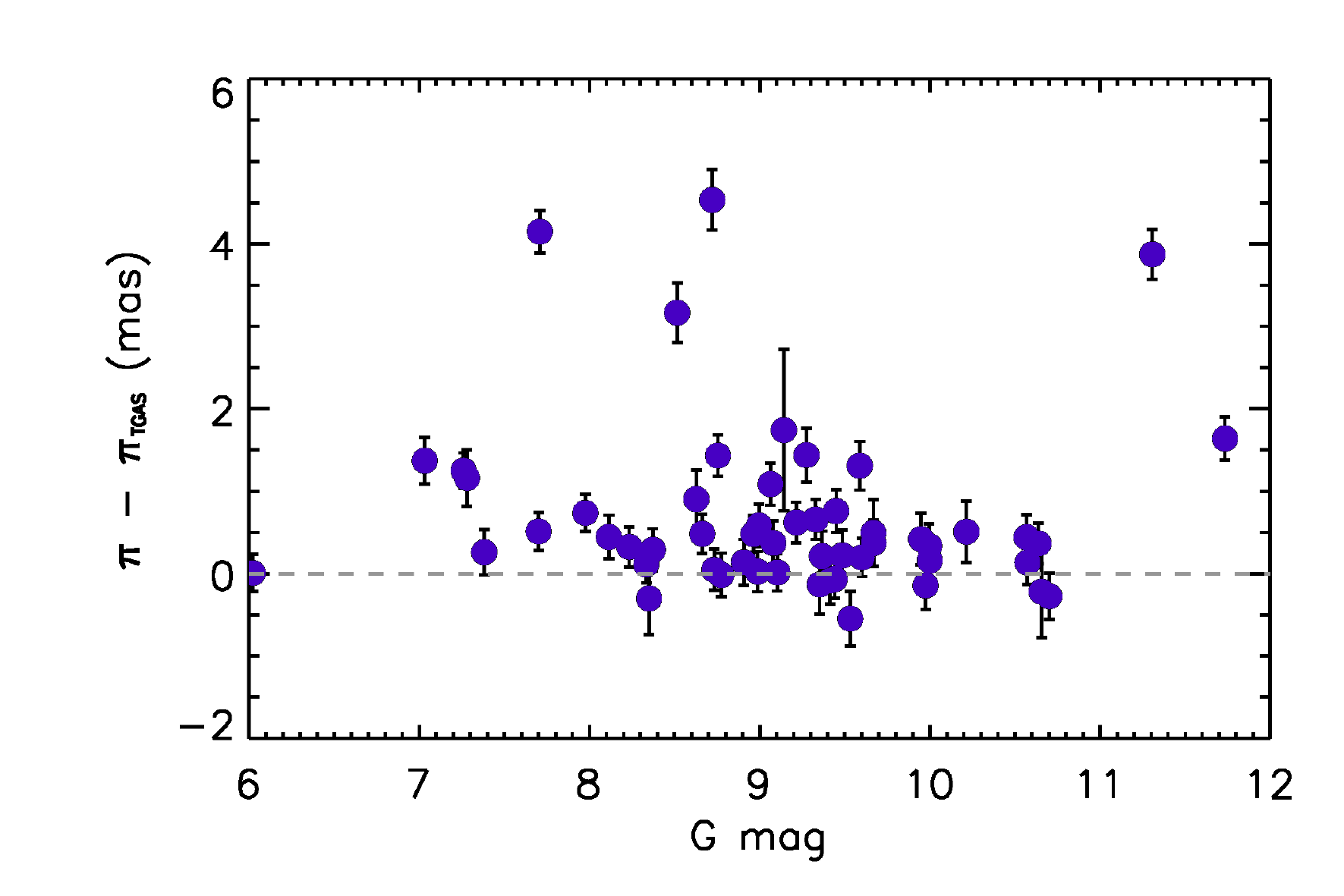}
    \caption{\label{fig:parallax}Comparison of the parallaxes derived in
this work with those from the TGAS solution.  Error bars are TGAS errors.}
\end{figure}

The TGAS catalogue of stellar parallaxes was recently made
available
through the first Gaia Data Release \citep{gaiamission}.  
A comparison of the parallaxes we derive and
these new values is shown in Figure~\ref{fig:parallax}. 
Apart from a few outliers there is an overall good agreement
between the independent methods. 
However, the parallaxes that we derive are systematically 
larger, with a mean difference of +0.7 mas, just over 2$\sigma$ the 
current uncertainties on the TGAS parallaxes.
The cause of this systematic difference is currently under investigation.
Nevertheless it is worthy to note that such comparisons of 
independent methods may allow us to reveal errors in the extinction measurements,
the bolometric corrections, the \teff\ 
or the underlying assumptions in the models which
compute the stellar luminosities.

\section{Trends in stellar properties \label{sec:stellartrends}}

Performing a homogenous analysis on a {large sample
allows us to check for trends in some stellar parameters,
and compare them to trends derived or established by other methods.
We performed this check for two parameters: the mixing-length
parameter and the stellar age.}

\subsection{The mixing-length parameter versus \teff\ and \logg}
The mixing-length parameter, $\alpha$, is
{usually} calibrated for a solar model and then applied to all 
models for a {set} range of masses and metallicities.  
However, several authors have shown that this {approach
is not correct} \citep{bonaca2012,creevey2012A}. 
{The values of $\alpha$ resulting from a GA offer}
an optimal approach to 
effectively test and subsequently constrain this parameter, since 
{the only}
assumption { is that $\alpha$ lies between 1 and 3.}

The distribution of $\alpha$ 
with \logg\ and \teff\ (colour-coded, see figure caption) is shown  
in Fig.~\ref{fig:poster_teffalpha}.
We see that for a given value of \logg, the 
{ value of $\alpha$ has an upper limit, which we can approximate by
$\alpha < 1.65 \log g - 4.75$, 
and this
is denoted by the dashed line in the figure.  
A regression analysis considering the model values of \logg, $\log$\teff\ and
[M/H]  yields: 
\begin{eqnarray}
\indent \alpha &=& 5.972778 + 0.636997 \log g \nonumber \\
  && - 1.799968 \log T_{\rm eff} + 0.040094 [{\rm M}/{\rm H}], \nonumber \label{eqn:alpha_regression}
\\
\end{eqnarray}
with a mean and rms of { the residual to the fit of --0.01 $\pm$ 0.15}.  
This equation yields a value of $\alpha =  2.03$ for the known solar properties.

\begin{figure}
    \includegraphics[width=0.45\textwidth]{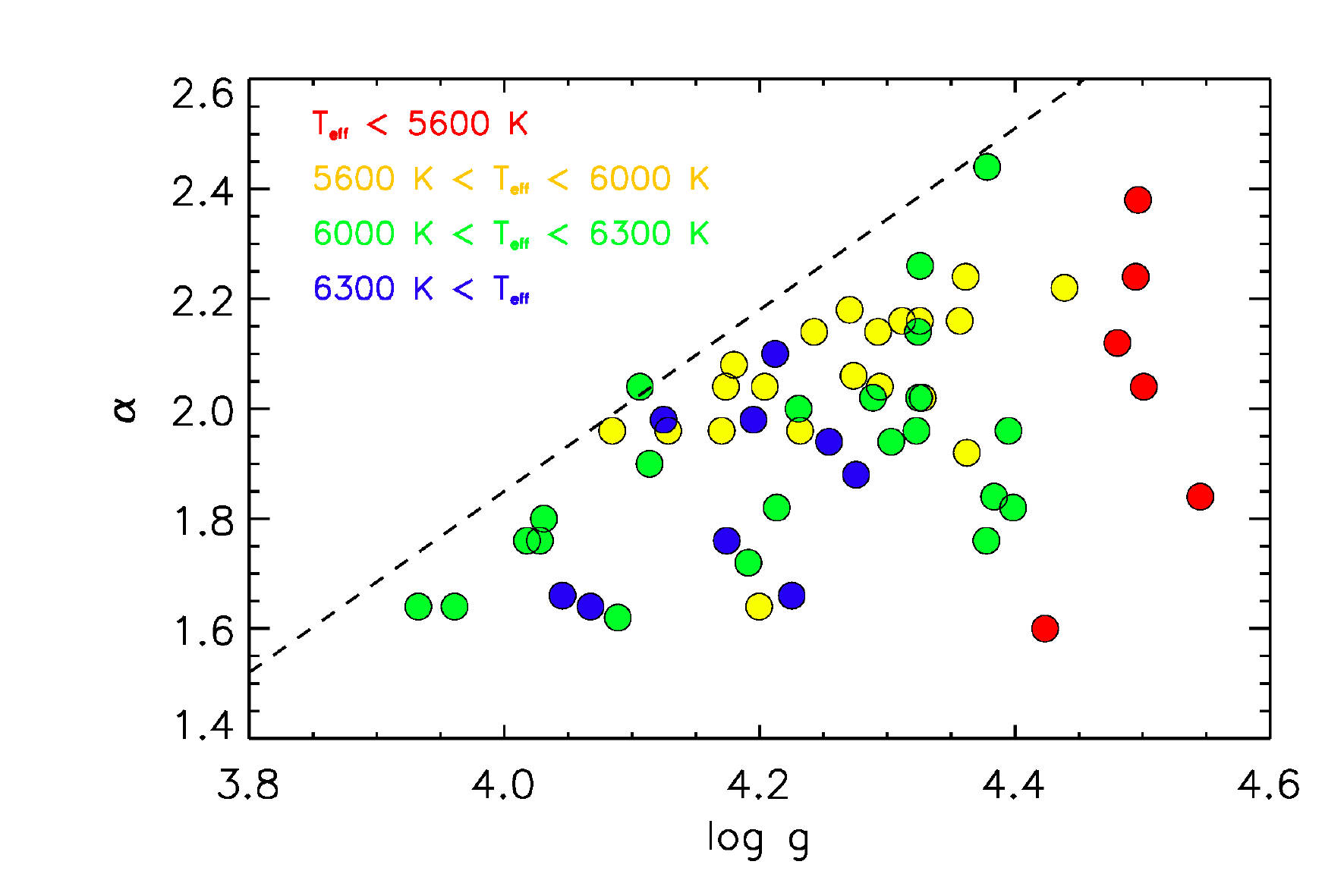}
    \includegraphics[width=0.45\textwidth]{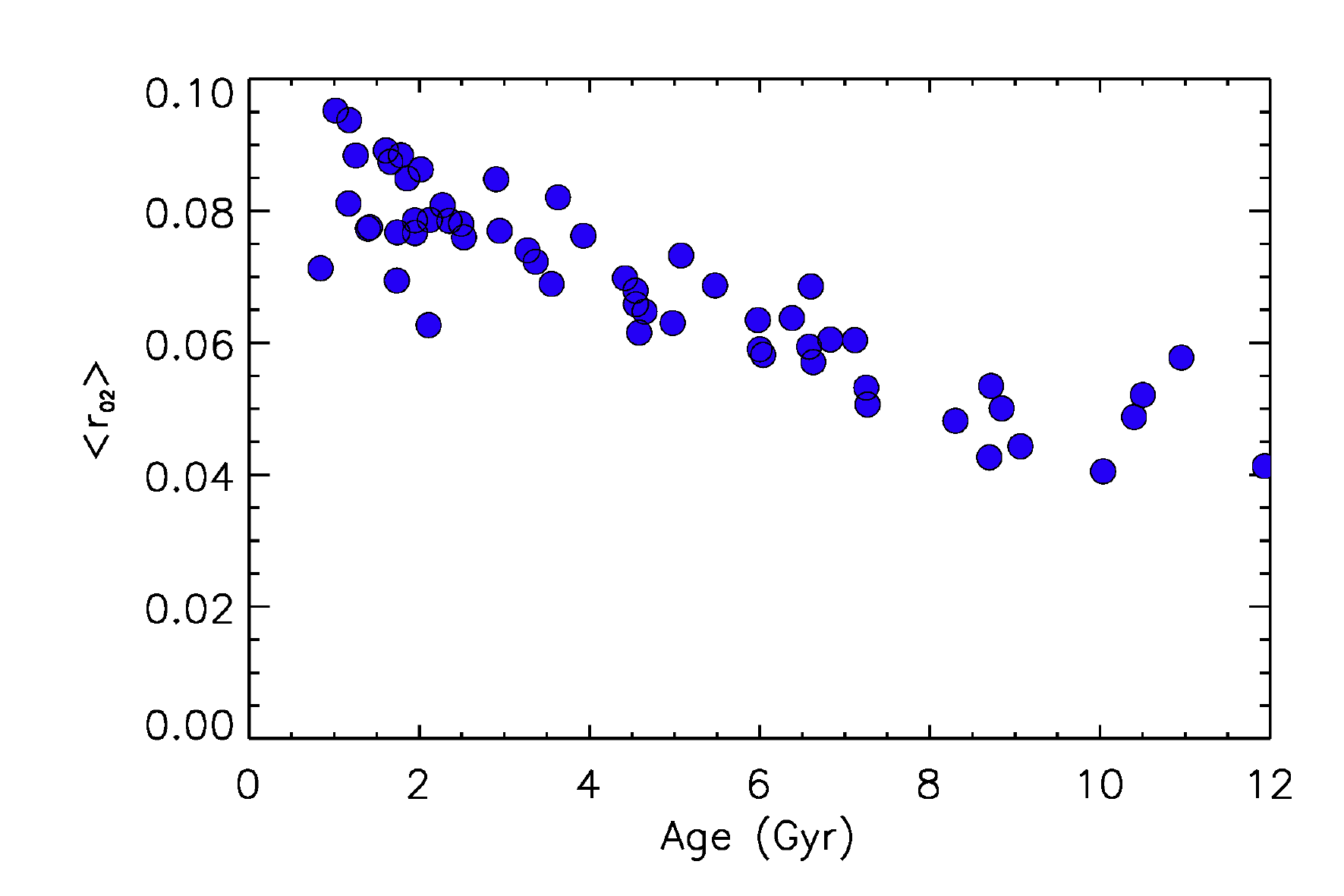}
    \caption{{\sl Top:} Distribution of the \logg\ and $\alpha$ for our sample.
        The colour coding is as follows: 
        red -- \teff\ $<$ 5600 K, 
        yellow -- 5600 K $<$ \teff\ $<$ 6000 K,
        green -- 6000 K $<$ \teff\ $<$ 6300 K,
        blue -- \teff\ $>$ 6300 K.
        {\sl Lower: }The distribution of age and  $\langle r02 \rangle$.  
    \label{fig:poster_teffalpha}}
\end{figure}

These results {agree in part} with those {derived} by 
\citep{magic2015}, using full 3D radiative hydrodynamic calculations for convective envelopes.
These authors also found that $\alpha$ increases with \logg\ and decreases with
\teff.
Our fit indicates a very small dependence on metallicity while their results
finds an opposite and more significant trend with this parameter.
Our sample, however, does not span a very large range in \mh, and the low 
coefficient is consistent with zero within the error bars.

\subsection{Age and $\langle r_{02} \rangle$}



The $r_{02}$ frequency separation ratios are effective at probing the gradients near the core of the star \citep{Roxburgh2003}.
As the core is most sensitive to nuclear processing, the $r_{02}$ are a diagnostic
of the evolutionary state of the star.  
{ Using} theoretical models \citep{lebMon2009} showed { a} relationship between
the mean value of $r_{02}$ and the { stellar} age.
{ That} relationship { was 
recently} used by \citep{appourchaux2015} to estimate the age of KIC~7510397 (HIP~93511).
Figure~\ref{fig:poster_teffalpha} shows the distribution of $ \langle r_{02} \rangle$  
versus the derived age for the sample of stars studied here.
{ A linear fit to these data leads to the following estimate of 
the stellar age, $\tau$, based on $ \langle r_{02} \rangle$}
\begin{equation}
    \tau = 17.910 - 193.918 \langle r_{02} \rangle,
    \label{eqn:r02}
\end{equation}
{ This is, of course, only valid for the range covered by our sample.
Note that when inserting the solar value of $\langle r_{02} \rangle = 0.068~\mu$Hz,
Eq.~\ref{eqn:r02} yields an age of 4.7 Gyr, in agreement with the Sun's
age as determined by other means.}

\section{Surface Effects}
\label{sec:surface}
The comparison of the observed 
oscillation frequencies from the Sun and other stars
with model frequencies calculated from 1D stellar models reveals
a systematic error in the models which increases with frequency.
These are known as {\it surface effects}, 
which arise from incomplete modelling of
the near-surface layers and the use of an adiabatic treatment on 
the stellar oscillation modes.   
Efforts to improve the stellar modelling is underway, but the 
application of improved models on a large scale is still out of 
reach.  
{To alleviate this problem, } several authors have advocated for the use 
of combination frequencies which are insensitive to this systematic
offset \citep{Roxburgh2003},
{hence the exclusive use of $r_{01}$ and $r_{02}$ in
the AMP~1.3 methodology}.
However, 
{since individual frequencies contain more information
than $r_{01}$ and $r_{02}$,}
some authors 
have derived simple prescriptions to 
mitigate the surface effect. 
One such parametrization is that of \citep{Kjeldsen2008} 
who suggest a simple correction to the individual frequencies 
$\delta\nu_{n,l}$ in the 
form {of a power law, namely}
\begin{equation}
    \delta\nu_{n,l} = a_0 \left ( \frac{\nu^{\rm obs}_{n,l}}{\nu_{\rm max}} \right )^{b}
        \label{eqn:kjeldsen}
\end{equation}
where $b = 4.82$ is a fixed value, {calibrated by a solar model}, 
$a_0$ is {computed from the differences between the} observed and model 
frequencies \citep{Metcalfe2009,Metcalfe2014}, 
{$\nu^{\rm obs}_{n,l}$ is the observed frequency $n,l$ mode
and  $\nu_{\rm max}$ is the frequency corresponding
to the highest amplitude mode, see \citep{Lund2016}}.  

The AMP~1.3 methodology uses exclusively
$r_{01}$ and $r_{02}$ as the seismic constraints, hence our results are 
insensitive
to the {surface effects}.  
{Using the model frequencies of the best-matched models,
we can  }
test how useful Eq.~\ref{eqn:kjeldsen} is for different 
stellar regimes
along the main sequence and early sub-giant phase.
The interest in testing this becomes apparent when we 
consider that for many stars we do not have a high enough
precision on the frequencies, or a sufficiently large range of
radial orders, to use $r_{01}$ and $r_{02}$ to effectively constrain 
the stellar modelling.  This is the case for some ground-based observations and
for some stars that will be observed by 
TESS \citep{tess-ricker2015}, and PLATO \citep{plato-rauer2014}, 
where a limited time series of only one to two months may
only be available.

In order to test where Eq.~\ref{eqn:kjeldsen} is useful, we calculated the residual between
the observed oscillation frequencies and the model frequencies 
corrected for the surface term (Eq.~\ref{eqn:kjeldsen}):   
$q_{n,l} = \nu^{\rm obs}_{n,l} - \nu^{\rm mod}_{n,l} + \delta\nu_{n,l},$.
Then we defined the metric  $Q$ as the median of the 
square root of the squared residuals:
\begin{equation}
    Q = \mathrm{median}\left \{\sqrt{q_{n,l}^2} \right \}, 
    \label{eqn:chit}
\end{equation}
for all $n$ and all $l$ defined in the region of 
$0.7 \le \nu_{n,l}^{\rm obs} / \nu_{\rm max} \le 1.3$.
We find values of $Q$ that vary between 0.3 and 10 and 
this variation is anti-correlated with $a_0$: as $a_0$ becomes more 
negative the match with the model becomes worse.   

We define a subsample of 44 stars with the best fits
to $r_{01}$ and $r_{02}$.  This subsample in shown in Fig.~\ref{fig:regions} 
as open circles for 
the observed values of \mlsep\ and \teff, and the 
derived values of mass and radius.  
In these figures the stars represented by the dark blue filled
circles have $Q \le 1.0$ and those represented by the light blue
filled circles have $1.0 < Q < 1.2$.  
It is evident from the figures that there exists regions of 
the parameter space where the surface correction proposed by \citep{Kjeldsen2008} 
adequately corrects for the surface term.
These regions are highlighted by the dashed lines, 
defined more conservatively here as
 $\log g > 4.2$, \teff\ $< 6250$ K, 
\mlsep\ $> 80 \mu$Hz and \numax\ $> 1700 \mu$Hz.
In physical properties this corresponds to a star with $R < 1.6$ \rsol,
$M < 1.3$ \msol, and $L < 2.5$ \lsol, 
with no apparent evidence of the age or the metallicity
playing any role.

\begin{figure*}
\centering
\includegraphics[width=0.45\textwidth,clip]{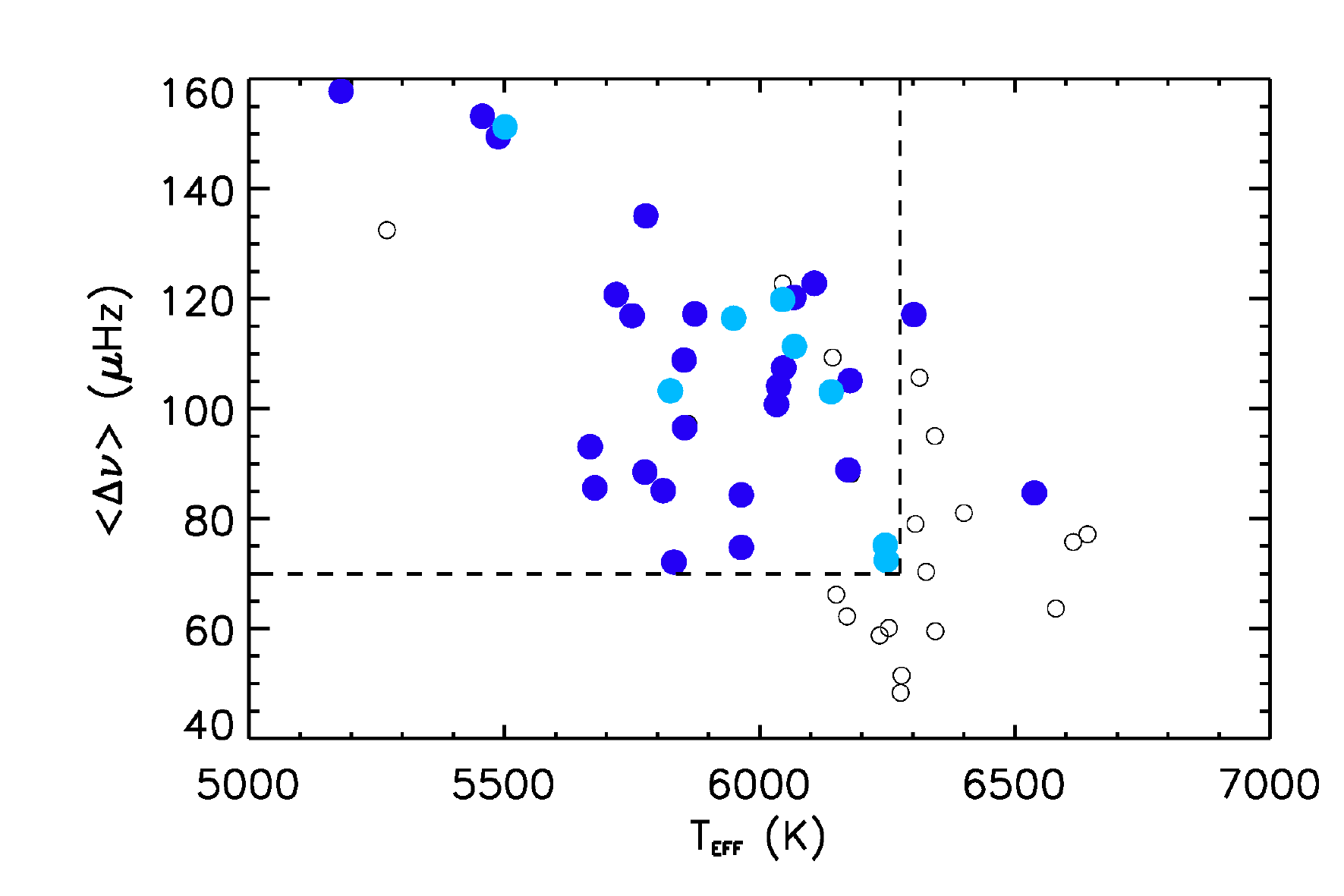}
\includegraphics[width=0.45\textwidth,clip]{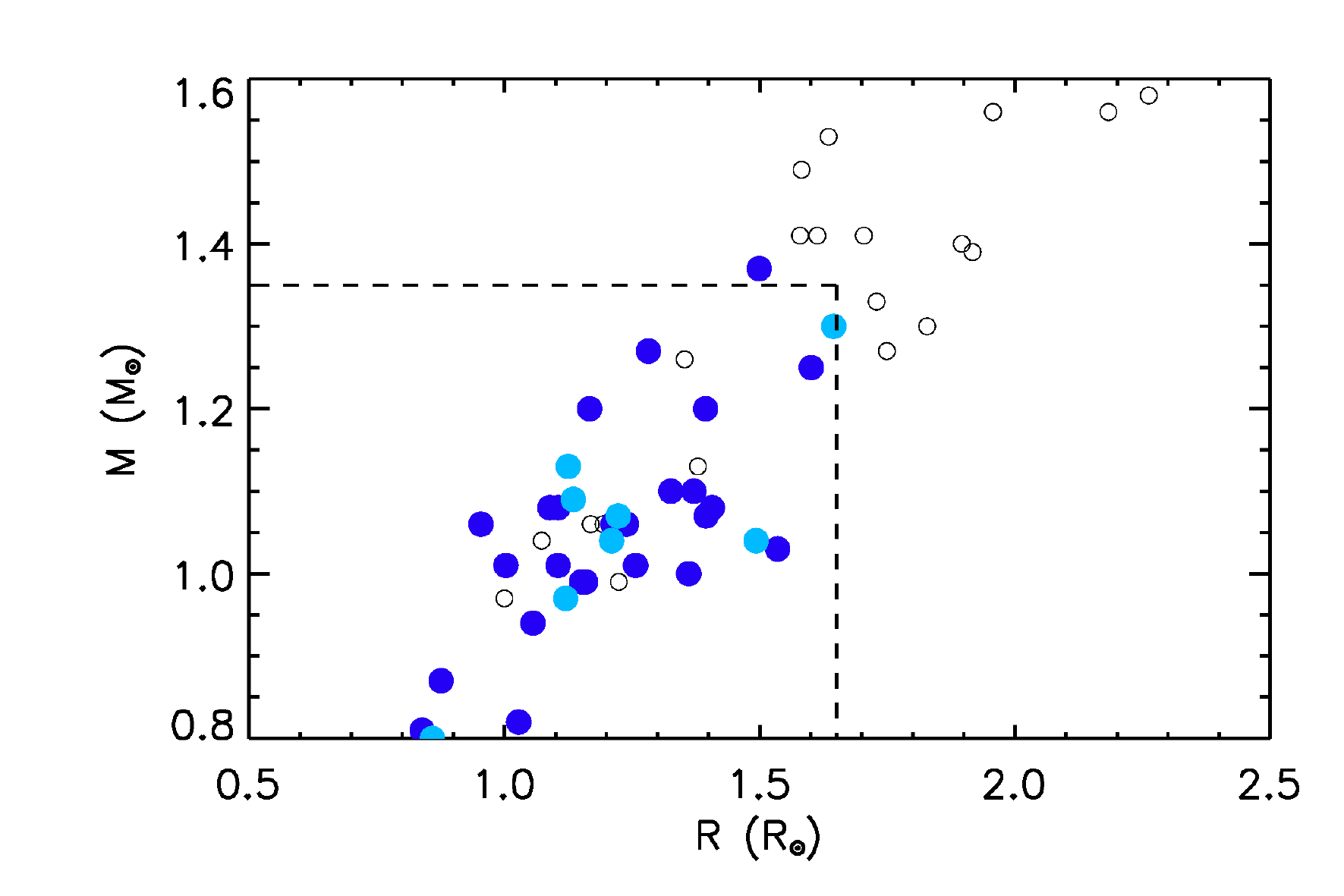}
    \caption{\label{fig:regions}Distribution of observed (left) and derived 
parameters (right) for a subsample of our stars (open circles).
The dark / light blue filled dots represent the stars with $Q \le 1.0$ / 1.2.
The prescription for correcting model frequencies is valid within the
region delimited by dashed lines.} 
\end{figure*}

\section{Conclusions}
\label{sec:conclusions}
In these proceedings we used the derived properties of 57 \emph{Kepler} 
stars presented in \citep{Creevey2016} to predict the star's
parallaxes, to derive an expression relating the mixing-length parameter
and the age to observed properties, and to explore the regions of 
parameters where the proposed surface correction to individual 
frequencies by \citep{Kjeldsen2008}
(Eq.~\ref{eqn:kjeldsen}) 
is useful.  For this latter point, the aim of defining valid regions
is to use the correction for data sets where the frequency precision
or the range of radial orders is not sufficient for $r_{01}$ and $r_{02}$ 
to constrain the stellar models.  
This will be the case for some stars that will be observed
in low ecliptic latitudes with TESS and for the {\it step-and-stare} 
pointing phases of the PLATO mission.

The parallaxes that we derived in this work were compared to the
parallaxes derived from the Tycho-Gaia solution.
We found in general a good agreement between our results.  
The mean differences between our results (ours -- theirs) is +0.7 mas
and the cause of this error is currently under study.  
Nevertheless we validate the new parallaxes from \citep{tgas2015}.
This comparison of parallaxes also highlights  
the ability of these external measurements to allow us to investigate the 
sources of systematic errors.  We believe that in providing a prior on
the luminosity, we may even be able to constrain the initial helium abundance
in the star.
We look forward to the wealth of forthcoming asteroseismic data on bright
nearby stars with TESS and PLATO, and the forthcoming Gaia parallaxes
with $\mu$as precision.
\bibliography{Creeveybib}

\begin{thebibliography}{25}

\bibitem{Lund2016}
M.N. {Lund}, V.~{Silva Aguirre}, G.R. {Davies}, W.J. {Chaplin},
  J.~{Christensen-Dalsgaard}, G.~{Houdek}, T.R. {White}, T.R. {Bedding}, W.H.
  {Ball}, D.~{Huber} et~al., \apj \textbf{835}, 172 (2017), \texttt{1612.00436}

\bibitem{Ramirez2009}
I.~{Ram{\'{\i}}rez}, J.~{Mel{\'e}ndez}, M.~{Asplund}, \aap \textbf{508}, L17
  (2009)

\bibitem{Buchhave2012}
L.A. {Buchhave}, D.W. {Latham}, A.~{Johansen}, M.~{Bizzarro}, G.~{Torres}, J.F.
  {Rowe}, N.M. {Batalha}, W.J. {Borucki}, E.~{Brugamyer}, C.~{Caldwell} et~al.,
  \nat \textbf{486}, 375 (2012)

\bibitem{Pinsonneault2012}
M.H. {Pinsonneault}, D.~{An}, J.~{Molenda-{\.Z}akowicz}, W.J. {Chaplin}, T.S.
  {Metcalfe}, H.~{Bruntt}, \apjs \textbf{199}, 30 (2012), \texttt{1110.4456}

\bibitem{Huber2013}
D.~{Huber}, W.J. {Chaplin}, J.~{Christensen-Dalsgaard}, R.L. {Gilliland},
  H.~{Kjeldsen}, L.A. {Buchhave}, D.A. {Fischer}, J.J. {Lissauer}, J.F. {Rowe},
  R.~{Sanchis-Ojeda} et~al., \apj \textbf{767}, 127 (2013), \texttt{1302.2624}

\bibitem{Chaplin2014}
W.J. {Chaplin}, S.~{Basu}, D.~{Huber}, A.~{Serenelli}, L.~{Casagrande},
  V.~{Silva Aguirre}, W.H. {Ball}, O.L. {Creevey}, L.~{Gizon}, R.~{Handberg}
  et~al., \apjs \textbf{210}, 1 (2014), \texttt{1310.4001}

\bibitem{Pinsonneault2014}
M.H. {Pinsonneault}, Y.~{Elsworth}, C.~{Epstein}, S.~{Hekker},
  S.~{M{\'e}sz{\'a}ros}, W.J. {Chaplin}, J.A. {Johnson}, R.A. {Garc{\'{\i}}a},
  J.~{Holtzman}, S.~{Mathur} et~al., \apjs \textbf{215}, 19 (2014),
  \texttt{1410.2503}

\bibitem{Roxburgh2003}
I.W. {Roxburgh}, S.V. {Vorontsov}, \aap \textbf{411}, 215 (2003)

\bibitem{Metcalfe2003}
T.S. {Metcalfe}, P.~{Charbonneau}, Journal of Computational Physics
  \textbf{185}, 176 (2003), \texttt{astro-ph/0208315}

\bibitem{Metcalfe2009}
T.S. {Metcalfe}, O.L. {Creevey}, J.~{Christensen-Dalsgaard}, \apj \textbf{699},
  373 (2009), \texttt{0903.0616}

\bibitem{Creevey2016}
O.~{Creevey}, T.S. {Metcalfe}, M.~{Schultheis}, D.~{Salabert}, M.~{Bazot},
  F.~{Thevenin}, S.~{Mathur}, H.~{Xu}, R.A. {Garcia}, ArXiv e-prints  (2016),
  \texttt{1612.08990}

\bibitem{tgas2015}
D.~{Michalik}, L.~{Lindegren}, D.~{Hobbs}, \aap \textbf{574}, A115 (2015),
  \texttt{1412.8770}

\bibitem{Kjeldsen2008}
H.~{Kjeldsen}, T.R. {Bedding}, J.~{Christensen-Dalsgaard}, \apjl \textbf{683},
  L175 (2008), \texttt{0807.1769}

\bibitem{schultheis2014}
M.~{Schultheis}, G.~{Zasowski}, C.~{Allende Prieto}, F.~{Anders}, R.L.
  {Beaton}, T.C. {Beers}, D.~{Bizyaev}, C.~{Chiappini}, P.M. {Frinchaboy}, A.E.
  {Garc{\'{\i}}a P{\'e}rez} et~al., \aj \textbf{148}, 24 (2014),
  \texttt{1405.2180}

\bibitem{2MASS}
M.F. {Skrutskie}, R.M. {Cutri}, R.~{Stiening}, M.D. {Weinberg}, S.~{Schneider},
  J.M. {Carpenter}, C.~{Beichman}, R.~{Capps}, T.~{Chester}, J.~{Elias} et~al.,
  \aj \textbf{131}, 1163 (2006)

\bibitem{marigo2008}
P.~{Marigo}, L.~{Girardi}, A.~{Bressan}, M.A.T. {Groenewegen}, L.~{Silva}, G.L.
  {Granato}, \aap \textbf{482}, 883 (2008), \texttt{0711.4922}

\bibitem{gaiamission}
{Gaia Collaboration}, T.~{Prusti}, J.H.J. {de Bruijne}, A.G.A. {Brown},
  A.~{Vallenari}, C.~{Babusiaux}, C.A.L. {Bailer-Jones}, U.~{Bastian},
  M.~{Biermann}, D.W. {Evans} et~al., \aap \textbf{595}, A1 (2016),
  \texttt{1609.04153}

\bibitem{bonaca2012}
A.~{Bonaca}, J.D. {Tanner}, S.~{Basu}, W.J. {Chaplin}, T.S. {Metcalfe},
  M.J.P.F.G. {Monteiro}, J.~{Ballot}, T.R. {Bedding}, A.~{Bonanno}, A.M.
  {Broomhall} et~al., \apjl \textbf{755}, L12 (2012), \texttt{1207.2765}

\bibitem{creevey2012A}
O.L. {Creevey}, F.~{Th{\'e}venin}, T.S. {Boyajian}, P.~{Kervella},
  A.~{Chiavassa}, L.~{Bigot}, A.~{M{\'e}rand}, U.~{Heiter}, P.~{Morel},
  B.~{Pichon} et~al., \aap \textbf{545}, A17 (2012), \texttt{1207.5954}

\bibitem{magic2015}
Z.~{Magic}, A.~{Weiss}, M.~{Asplund}, \aap \textbf{573}, A89 (2015),
  \texttt{1403.1062}

\bibitem{lebMon2009}
Y.~{Lebreton}, J.~{Montalb{\'a}n}, \emph{{Stellar ages from asteroseismology}},
  in \emph{The Ages of Stars}, edited by E.E. {Mamajek}, D.R. {Soderblom},
  R.F.G. {Wyse} (2009), Vol. 258 of \emph{IAU Symposium}, pp. 419--430,
  \texttt{0811.2908}

\bibitem{appourchaux2015}
T.~{Appourchaux}, H.M. {Antia}, W.~{Ball}, O.~{Creevey}, Y.~{Lebreton},
  K.~{Verma}, S.~{Vorontsov}, T.L. {Campante}, G.R. {Davies}, P.~{Gaulme}
  et~al., \aap \textbf{582}, A25 (2015)

\bibitem{Metcalfe2014}
T.S. {Metcalfe}, O.L. {Creevey}, G.~{Do{\u g}an}, S.~{Mathur}, H.~{Xu}, T.R.
  {Bedding}, W.J. {Chaplin}, J.~{Christensen-Dalsgaard}, C.~{Karoff},
  R.~{Trampedach} et~al., \apjs \textbf{214}, 27 (2014), \texttt{1402.3614}

\bibitem{tess-ricker2015}
G.R. {Ricker}, J.N. {Winn}, R.~{Vanderspek}, D.W. {Latham}, G.{\'A}. {Bakos},
  J.L. {Bean}, Z.K. {Berta-Thompson}, T.M. {Brown}, L.~{Buchhave}, N.R.
  {Butler} et~al., Journal of Astronomical Telescopes, Instruments, and Systems
  \textbf{1}, 014003 (2015)

\bibitem{plato-rauer2014}
H.~{Rauer}, C.~{Catala}, C.~{Aerts}, T.~{Appourchaux}, W.~{Benz},
  A.~{Brandeker}, J.~{Christensen-Dalsgaard}, M.~{Deleuil}, L.~{Gizon}, M.J.
  {Goupil} et~al., Experimental Astronomy \textbf{38}, 249 (2014),
  \texttt{1310.0696}

\end{thebibliography}
\end{document}